\documentclass[pdflatex,sn-mathphys-num,referee]{sn-jnl}

\usepackage{graphicx}%
\usepackage{multirow}%
\usepackage{amsmath,amssymb,amsfonts}%
\usepackage{amsthm}%
\usepackage{mathrsfs}%
\usepackage[title]{appendix}%
\usepackage{xcolor}%
\usepackage{textcomp}%
\usepackage{manyfoot}%
\usepackage{booktabs}%
\usepackage{algorithm}%
\usepackage{algorithmicx}%
\usepackage{algpseudocode}%
\usepackage{listings}%
\usepackage{hyperref}%

\theoremstyle{thmstyleone}%
\theoremstyle{thmstyletwo}%

\theoremstyle{thmstylethree}%

\begin{document}

\title[Mass regulates the emerging timescale of young star clusters]{Mass regulates the emerging timescale of young star clusters}


\author*[1]{\fnm{Alex} \sur{Pedrini}}\email{alex.pedrini@astro.su.se}

\author*[1]{\fnm{Angela} \sur{Adamo}}\email{angela.adamo@astro.su.se}

\author[2]{\fnm{Daniela} \sur{Calzetti}}
\equalcont{These authors contributed equally to this work.}

\author[1]{\fnm{Arjan} \sur{Bik}}
\equalcont{These authors contributed equally to this work.}

\author[3]{\fnm{Thomas J.} \sur{Haworth}}
\equalcont{These authors contributed equally to this work.}

\author[4]{\fnm{Bruce G.} \sur{Elmegreen}}
\equalcont{These authors contributed equally to this work.}

\author[5]{\fnm{Mark R.} \sur{Krumholz}}
\equalcont{These authors contributed equally to this work.}

\author[6]{\fnm{Sean T.} \sur{Linden}}
\equalcont{These authors contributed equally to this work.}

\author[2]{\fnm{Benjamin} \sur{Gregg}}
\equalcont{These authors contributed equally to this work.}

\author[1]{\fnm{Helena} \sur{Faustino Vieira}}
\equalcont{These authors contributed equally to this work.}

\author[7]{\fnm{Varun} \sur{Bajaj}}
\equalcont{These authors contributed equally to this work.}

\author[7]{\fnm{Jenna E.} \sur{Ryon}}
\equalcont{These authors contributed equally to this work.}

\author[8]{\fnm{Ahmad A.} \sur{Ali}}
\equalcont{These authors contributed equally to this work.}

\author[9]{\fnm{Eric P.} \sur{Andersson}}
\equalcont{These authors contributed equally to this work.}

\author[1]{\fnm{Giacomo} \sur{Bortolini}}
\equalcont{These authors contributed equally to this work.}

\author[10]{\fnm{Michele} \sur{Cignoni}}
\equalcont{These authors contributed equally to this work.}

\author[11]{\fnm{Ana} \sur{Duarte-Cabral}}
\equalcont{These authors contributed equally to this work.}

\author[12,13,14]{\fnm{Kathryn} \sur{Grasha}}
\equalcont{These authors contributed equally to this work.}

\author[15,19]{\fnm{Natalia} \sur{Lahén}}
\equalcont{These authors contributed equally to this work.}

\author[16]{\fnm{Thomas} \sur{S.-Y. Lai}}
\equalcont{These authors contributed equally to this work.}

\author[2]{\fnm{Drew} \sur{Lapeer}}
\equalcont{These authors contributed equally to this work.}

\author[17]{\fnm{Matteo} \sur{Messa}}
\equalcont{These authors contributed equally to this work.}

\author[1]{\fnm{Göran} \sur{Östlin}}
\equalcont{These authors contributed equally to this work.}

\author[18]{\fnm{Elena} \sur{Sabbi}}
\equalcont{These authors contributed equally to this work.}

\author[7]{\fnm{Linda} \sur{J. Smith}}
\equalcont{These authors contributed equally to this work.}

\author[17]{\fnm{Monica} \sur{Tosi}}
\equalcont{These authors contributed equally to this work.}

\affil*[1]{\orgdiv{Department of Astronomy}, \orgname{Stockholm University \& Oskar Klein Center}, \orgaddress{\street{AlbaNova University center}, \city{Stockholm}, \postcode{10691}, \country{Sweden}}}

\affil[2]{\orgdiv{Department of Astronomy}, \orgname{University of Massachusetts}, \orgaddress{\street{710 North Pleasant Street}, \city{Amherst}, \postcode{01003}, \state{MA}, \country{USA}}}

\affil[3]{\orgdiv{Astronomy Unit, School of Physics and Astronomy}, \orgname{Queen Mary University of London}, \orgaddress{\street{Mile End}, \city{London}, \postcode{E1 4NS}, \country{UK}}}

\affil[4]{\orgaddress{\city{Katonah}, \postcode{10536}, \state{NY}, \country{USA}}}

\affil[5]{\orgdiv{Research School of Astronomy \& Astrophysics}, \orgname{Australian National University}, \orgaddress{\street{233 Mt Stromlo Rd}, \city{Stromlo}, \postcode{ACT 2611}, \country{Australia}}}

\affil[6]{\orgdiv{Steward Observatory}, \orgname{University of Arizona}, \orgaddress{\street{933 N Cherry Avenue}, \city{Tucson}, \postcode{85721}, \state{AZ}, \country{USA}}}

\affil[7]{\orgname{Space Telescope Science Institute}, \orgaddress{\street{3700 San Martin Drive}, \city{Baltimore}, \postcode{21218}, \state{MD}, \country{USA}}}

\affil[8]{\orgdiv{I. Physikalisches Institut}, \orgname{Universit\"{a}t zu K\"{o}ln}, \orgaddress{{Z\"{u}lpicher Str. 77}, \city{K\"{o}ln}, \postcode{50937}, \country{Germany}}}

\affil[9]{\orgdiv{Department of Astrophysics}, \orgname{American Museum of Natural History}, \orgaddress{\street{200 Central Park West}, \city{New York}, \postcode{10024}, \state{NY}, \country{USA}}}

\affil[10]{\orgdiv{Dipartimento di Fisica}, \orgname{Università di Pisa}, \orgaddress{\street{Largo Bruno Pontecorvo 3}, \city{Pisa}, \postcode{56127}, \country{Italy}}}

\affil[11]{\orgdiv{Cardiff Hub for Astrophysics Research and Technology (CHART), School of Physics \& Astronomy}, \orgname{Cardiff University}, \orgaddress{\street{The Parade}, \city{Cardiff}, \postcode{CF24 3AA}, \country{UK}}}

\affil[12]{\orgdiv{Research School of Astronomy \& Astrophysics}, \orgname{Australian National University}, \city{Canberra}, \postcode{ACT 2611}, \country{Australia}}

\affil[13]{\orgdiv{ARC Centre of Excellence for All Sky Astrophysics in 3 Dimensions (ASTRO 3D)}, \country{Australia}}

\affil[14]{ARC DECRA Fellow}

\affil[15]{\orgname{Max-Planck-Institut für Astrophysik}, \orgaddress{\street{Karl-Schwarzschild-Str. 1}, \city{Garching}, \postcode{D-85748}, \country{Germany}}}

\affil[16]{\orgdiv{IPAC},\orgname{California Institute of Technology}, \orgaddress{\street{1200 E. California Blvd.}, \city{Pasadena}, \postcode{91125},\state{CA}, \country{USA}}}

\affil[17]{\orgdiv{INAF – OAS},\orgname{Osservatorio di Astrofisica e Scienza dello Spazio di Bologna}, \orgaddress{\street{via Gobetti 93/3}, \city{Bologna}, \postcode{I-40129}, \country{Italy}}}

\affil[18]{\orgname{Gemini Observatory/NSFs NOIRLab}, \orgaddress{\street{950 N. Cherry Ave.}, \city{Tucson}, \postcode{85719},\state{AZ}, \country{USA}}}

\affil[19]{\orgname{Zentrum f\"ur Astronomie der Universit\"at Heidelberg, Astronomisches
Rechen-Institut}, \orgaddress{\street{M\"onchhofstr. 12-14}, \city{Heidelberg}, \postcode{D-69120}, \country{Germany}}}


\maketitle

{\bf

Quantifying the timescales of star cluster emergence from their natal clouds remains one of the main challenges in our understanding of the star formation process.
These timescales are fundamental measurements of  the star formation cycle within galaxies, yet they are difficult to constrain due to the complex interplay between stellar feedback and star formation across a vast range of physical scales. Here we present HST and JWST observations of thousands of young star clusters in four nearby galaxies (M51, M83, NGC 628, and NGC 4449). A substantial fraction of these clusters are still embedded within their natal gas and remain invisible at optical wavelengths. We constrain their emergence process by measuring the timescales required to disperse the surrounding material. We find a strong correlation between dispersal timescale and cluster stellar mass, with massive clusters emerging more rapidly than their lower mass counterparts. This is a critical constraint on simulations of star formation and stellar feedback, which struggle to fully reproduce the formation and emergence of star clusters. Our results emphasize the central role of massive clusters in driving the escape of ionizing radiation into the galactic medium. Finally, they impose important limitations to the time available for planet formation in massive cluster environments where disks get exposed to UV irradiation and further gas infall is shut off.

}

\section{Introduction}

Star formation is generally understood to occur through the gravitational collapse of high-density regions in giant molecular clouds, leading to the formation of clustered stellar systems \cite{Lada03}. In star clusters and associations, massive stars ($M >15$ M$_{\odot}$) are responsible for the vast majority of energy and momentum released into the surrounding medium through a combination of processes, commonly referred to as stellar feedback which leads to the destruction of their natal cloud (photoionization, stellar winds, radiation pressure, and supernovae). Stellar feedback is thus one of the main processes believed to explain the observed inefficiency of star formation in galaxies, where only fractions of the available gas will ever form stars \cite[e.g.,][]{Krumholz19}. However, the timescale by which feedback drives cloud dispersal and its dependence on both star cluster properties and the surrounding environment remain poorly understood.

In recent years, numerical simulations have made significant progress in incorporating the key physical processes necessary to study cluster formation and evolution \cite[][among many others]{Grudic22, Ali23, Polak24a, Andersson24, Menon25, Lahen25}. This has led to major advances in modeling observations of young star clusters (YSCs) and in understanding the detailed role of feedback in shaping their evolution and emergence. Nevertheless, many aspects of the star formation process across different physical scales remain uncertain, and even the most sophisticated galaxy simulations still struggle to fully reproduce the life cycle of star clusters.

From an observational perspective, estimating the emerging timescales of YSCs in both the Milky Way and the Local Volume ($d<10$ Mpc) has been challenging. In the Milky Way, line-of-sight confusion, obscuration, and the intrinsic complexity of individual systems hinder a statistical and general interpretation. On the other hand, until recently, extragalactic observations have only been able to resolve tracers of star formation and cold molecular gas (e.g., H$\alpha$ and CO emission) at the physical scales of star-forming complexes \citep[$\sim 100$ pc,][]{Grasha18,Chevance22,Kim25, Ramambason25}, but not down to the parsec scales necessary to detect single molecular clouds and star clusters.

The excellent sensitivity and resolution of the Hubble Space Telescope (HST) and the James Webb Space Telescope (JWST) enable us to map populations of YSCs in nearby galaxies and their associated star-forming regions, thus providing constraints on their emerging sequence. With the advent of JWST, several surveys have been characterizing populations of dusty emerging YSCs (eYSCs) in nearby galaxies \citep{Sun24,Linden24,Rodriguez25,Knutas25,Whitmore25, Henny25}. These clusters show prominent ionized hydrogen and 3.3 $\mu$m polycyclic aromatic hydrocarbon (PAH) emission, which trace their evolutionary stages \citep{Pedrini24,Whitmore25,Knutas25}.

\begin{figure}
    \centering
    \includegraphics[width=\textwidth]{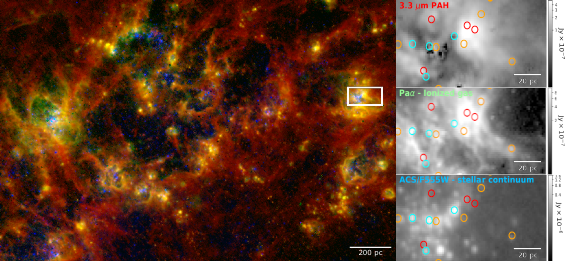}
    \caption{A kpc-scale star-forming complex in M51. {\it Left panel:} RGB composite image of the complex observed with HST and JWST. The blue channel traces stellar optical emission from the HST ACS/F555W filter, the green channel shows ionized emission from the Br$\alpha$ recombination line (JWST F405N, continuum-subtracted), and the red channel displays 3.3 $\mu$m PAH emission observed with JWST F335M (continuum-subtracted). The white box indicates the region shown in the right panels.
    {\it Right panels:} Monochromatic maps of the zoomed-in region highlighted by the white box. From top to bottom: 3.3 $\mu$m PAH, Pa$\alpha$, and ACS/F555W. Positions of eYSCs and optically detected YSCs are marked with red, orange, and cyan circles for eYSCI, eYSCII, and optical YSCs, respectively. In M51, photometry of these clusters has been performed at 5.8 pc scale. The eYSC and optical YSC classification is described in the Results.
    }
    \label{fig:rgb}
\end{figure}

In Figure~\ref{fig:rgb}, we show a representative example of complex star-forming regions in M51 observed with HST and JWST. 
The gas in the proximity of YSCs ($\sim 1-7$ Myr old) where massive stars are located is ionized by very energetic photons ($E > 13.6$ eV) and emits in the Br$\alpha$-4.05~$\mu$m recombination line (continuum subtracted F405N filter, green channel in Figure~\ref{fig:rgb}). After the ionization front, far-UV photons permeate the medium, dissociating molecular hydrogen and exciting PAH molecules \cite{Allamandola89, Tielens08}, which dominate the mid-infrared spectra of photo-dissociation regions (PDRs). In particular, the 3.3 $\mu$m PAH emission (continuum subtracted F335M filter, red channel in Figure~\ref{fig:rgb}) traces PDRs at the same physical resolution of the hydrogen recombination lines. 

In this work we take advantage of extensive JWST and HST multiwavelength campaigns to constraint the emerging timescales of star clusters from a deeply embedded phase to a optically exposed state at physical scales of 4--to--8 pc. We use the morphology of H\,\textsc{ii} regions and PDRs surrounding eYSCs to establish an evolutionary sequence and constrain the relative ages of the different emergence stages as a function of the stellar masses of the clusters.

\section{Results} 

We used JWST/NIRCam observations to identify eYSCs at 4-8 parsec scale in four nearby galaxies among the Feedback in Emerging extrAgalactic Star clusTers (FEAST, PI A. Adamo, \#1783) sample: M51, M83, NGC 628, NGC 4449. These galaxies have been selected to span various galactic environments and metallicities. Furthermore, we used archival HST imaging to identify optical YSCs (oYSCs) at the same spatial scale of the eYSCs.
We then define three classes of YSCs based on the following identification criteria:   
\begin{itemize}
\item eYSCI: eYSCs identified as compact and bright sources in the Pa$\alpha$/Br$\alpha$ emission lines and in the 3.3 $\mu$m PAH feature (red circles in Figure~\ref{fig:rgb}). These objects still show compact H\,\textsc{ii} regions and PDRs associated with the stellar component.
\item eYSCII: in this case, we do not observe a compact 3.3 $\mu$m PAH feature, even if emission is observed. However, we still detect compact and bright Pa$\alpha$/Br$\alpha$ emission (orange circles in Figure~\ref{fig:rgb}).
\item  oYSC: optically selected star clusters (bright sources in V band, ACS/F555W) younger than 10 Myr (blue circles in Figure~\ref{fig:rgb}) based on spectral energy distribution (SED) fitting outputs, introduced below. 
\end{itemize}

We refer to the Methods section for a complete description of all the analysis steps leading to the creation of final catalogues. In general, source extraction and identification was followed by multi band photometry across all available HST and JWST bands. SED fitting was then performed on the selected clusters using {\tt CIGALE} (Code Investigating GALaxy Emission \citep{Boquien19}), yielding physical parameters for the entire population of eYSCs and oYSCs, including age, stellar mass and extinction. Both the photometric measurements and the SED fitting procedure are described in details in \cite{Pedrini25} and are summarized in the Methods. 
For each class and galaxy, we report the total number of selected cluster candidates in each category in Extended Data Table~\ref{tab:numbers}.

\begin{figure}
    \centering
    \includegraphics[width=0.65\linewidth]{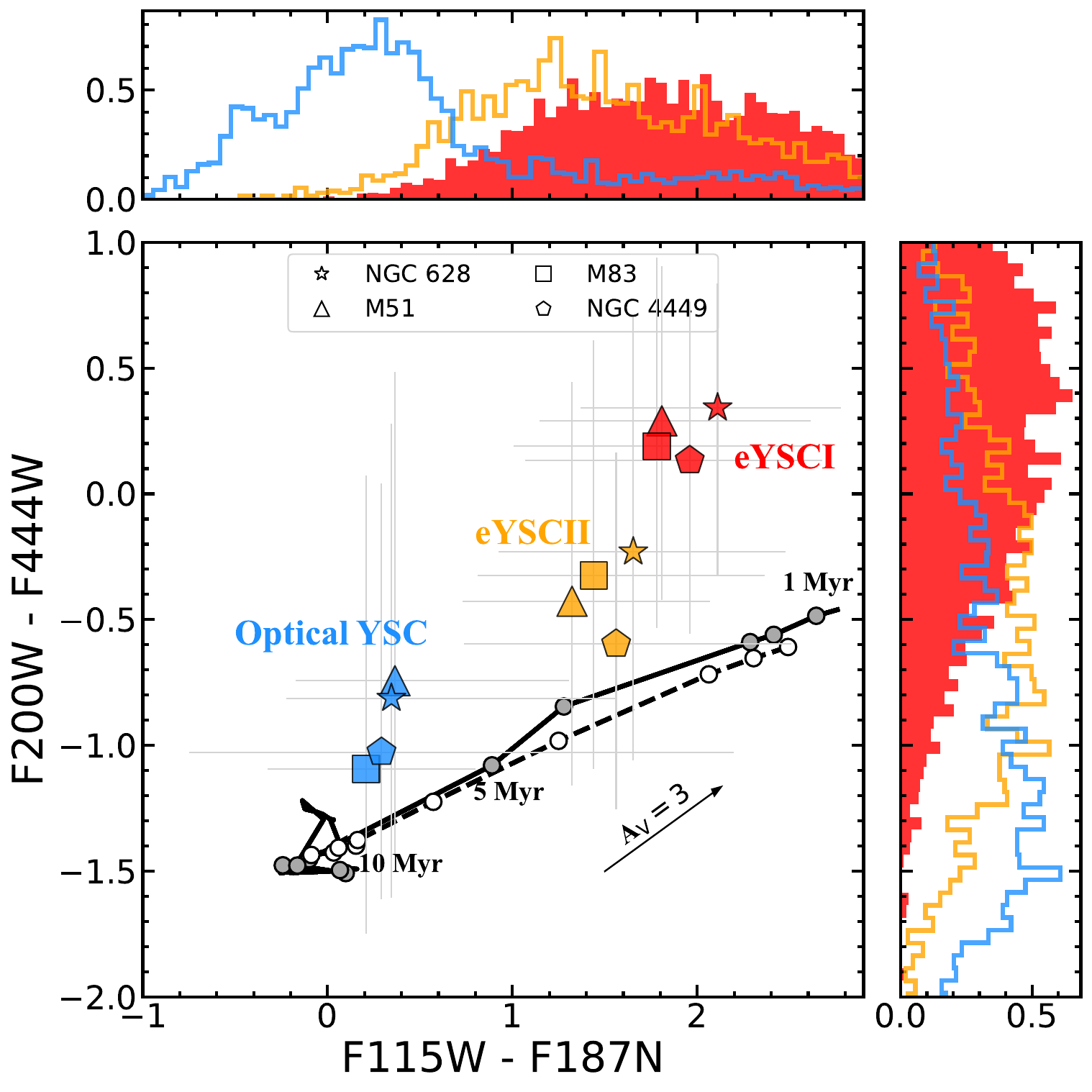}
    \hfill
    \includegraphics[width=0.65\linewidth]{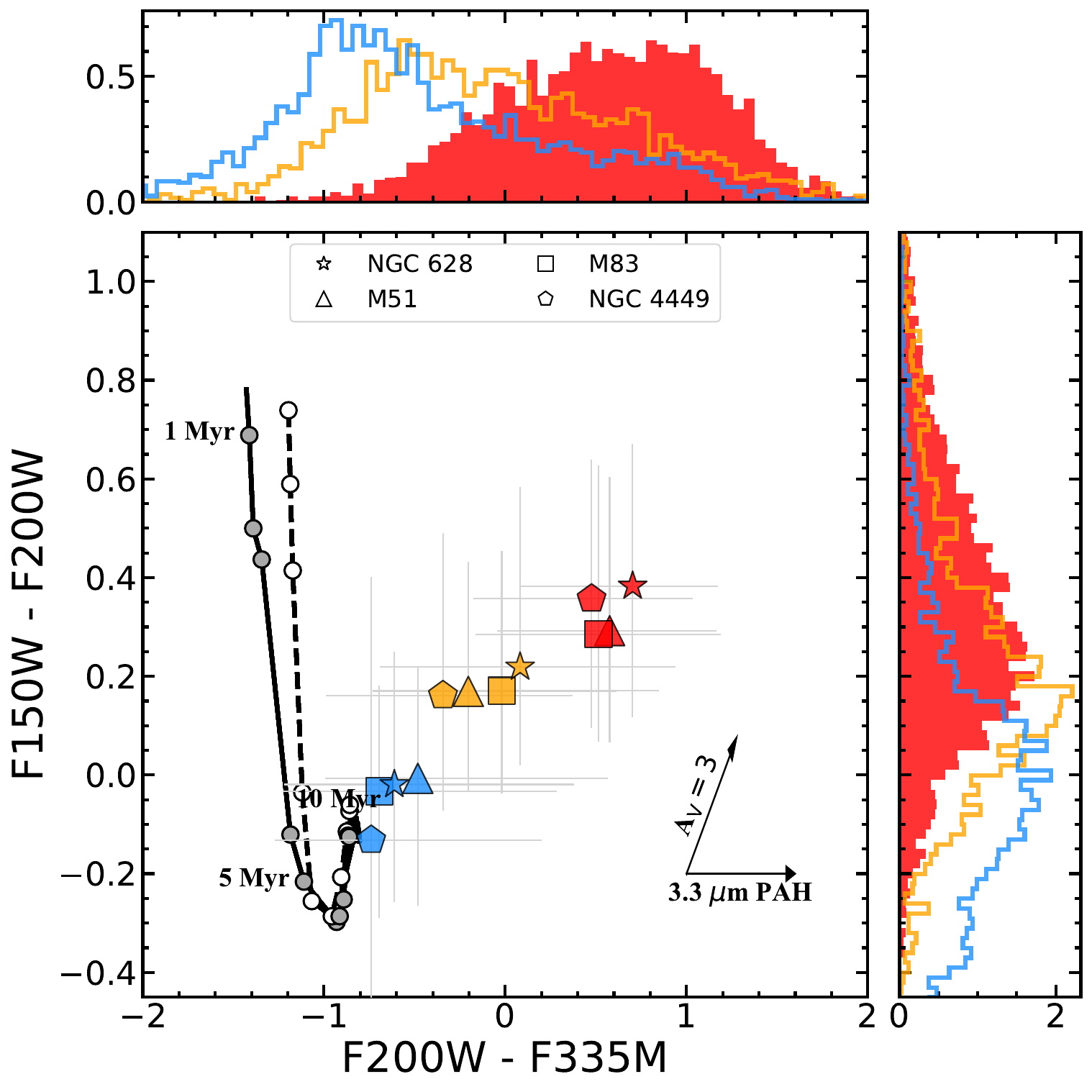}
    \caption{{\it Top:} NIR color–color diagram of the eYSC and oYSCs populations in the four FEAST galaxies. Units are AB mag. Redder colors along the y-axis trace hot dust emission through the F444W filter, while along the x-axis the F187N filter shifts younger objects to redder colors owing to bright Pa$\alpha$ emission. Symbols show the median values for the four galaxies, color-coded by cluster class, with the color scheme matching that in Figure~\ref{fig:rgb}. Errorbars represent the 16-84 percentiles of each distribution. Histograms display the combined distributions, also color-coded by cluster class. The evolutionary tracks are taken from the {\tt yggdrasil} single stellar population synthesis code (solid) and {\tt CIGALE} (dashed), with nebular emission and no dust absorption or emission, as described in the text. The extinction vector is from \cite{Fahrion23}. {\it Bottom:} Same as above, but with a different color combination. Here, the y-axis (F150W–F200W) serves as a proxy for young star cluster age, as indicated by the evolutionary track. The x-axis traces bright emission from the 3.3 $\mu$m PAH feature, displayed as an excess in the F335M filter.}
    \label{fig:NIRcolors}
\end{figure}

In Figure~\ref{fig:NIRcolors}, we present the near-infrared (NIR) colors for eYSCs and oYSCs. For each galaxy and emergence class, we show the median color values (symbols, eYSCI in red, eYSCII in orange, and oYSCs in blue) and the 16-84 percentiles of each sample distribution (bars). We also display histograms of the combined distributions for the four galaxies at the top and right side of each plot. In both panels, the median points and the histograms reveal three distinct populations corresponding to the three classes. The solid and dashed evolutionary tracks are extracted from \texttt{yggdrasil} \citep{Zackrisson11} and {\tt CIGALE}, respectively, and represent single stellar populations based on Starburst99/Padova-AGB \citep{Vazquez05} models  vs. BC03 \citep{Bruzual03} stellar tracks, respectively. Both models include nebular emission treatment \citep{Ferland13} assuming a covering fraction of 0.5, solar metallicity, no extinction and no dust emission. Although \texttt{CIGALE} includes dust and PAH emission in our standard setup (see Methods), these components were intentionally switched off for this figure to produce dust-free reference tracks, consistent with the assumptions adopted for the \texttt{yggdrasil} models.

In all galaxies investigated in this work, the median colors of the three classes show the same trend in both panels. The youngest class (eYSCI, red symbols) is associated with the largest excess in Pa$\alpha$ and in the redder F444W filter, tracers of both ionized gas and emission from hot dust (top panel of Figure ~\ref{fig:NIRcolors}), as well as in 3.3 $\mu$m emission  and dust extinction (bottom panel of Figure ~\ref{fig:NIRcolors}). The eYSCII (orange symbols) are an intermediate stage of emergence. They still present significant excess emission from hot dust and PAHs, causing their colors to deviate from the dust-free model tracks, even when accounting for dust extinction (A$_{\rm V}$ vector). In contrast, oYSCs populate regions of the NIR diagrams consistent with older ages (7 - 10 Myr), moderate dust extinction, and no or minimal dust emission. 
The NIR colors of these three different classes of star clusters therefore reveal an evolutionary sequence observed in all the studied galaxies: eYSCI $\rightarrow$ eYSCII $\rightarrow$ oYSCs, which we later use to quantify the timescales of cluster emergence.

In general, the {\tt CIGALE} results reinforce the observed evolutionary sequence of Figure~\ref{fig:NIRcolors}, with star clusters becoming older and less reddened when moving from eYSCI to oYSCs \citep{Knutas25}. However, as shown in a recent study \citep{Pedrini25}, systematic uncertainties in the absolute ages recovered from SED fitting affect the timescales analysis. We therefore use relative numbers of clusters in the different emergence stages to derive the emerging timescales. We define the {\it emerging} timescale $\tau_{\rm TOT}$ as the period required for clusters with associated H\,\textsc{ii} regions (eYSCI and eYSCII) to evolve into exposed oYSCs, e.g. total gas removal. This timescale is given by the following:
\begin{equation}
    \tau_{\rm TOT} = \frac{\#{\rm eYSCI} + \#{\rm eYSCII}}{\#{\rm eYSCI} + \#{\rm eYSCII} + \#{\rm oYSC }} \times 10 \, {\rm Myr}
\label{eq:total}
\end{equation}
Where the symbol \# defines the number of clusters in each class. We note that eYSCs overlapping in position with optically identified sources are counted only as oYSCs (see Methods).
We also define the PDR timescale $\tau_{\rm PDR}$ as the period over which eYSCI lose their compact PDRs, traced by the 3.3 $\mu$m PAH emission. This would correspond to the most embedded phase and is estimated as:
\begin{equation}
    \tau_{\rm PDR} = \frac{\#{\rm eYSCI}}{\#{\rm eYSCI} + \#{\rm eYSCII} + \#{\rm oYSC }} \times 10 \, {\rm Myr}
\label{eq:pdr}
\end{equation}

Details of this method and the derivation of these timescales are presented in the Methods. In defining these timescales, our starting point is the onset of photoionization by massive stars of the surrounding neutral hydrogen, traced by Pa$\alpha$ emission. The earlier, deeply embedded phase, during which clusters remain obscured and the ionizing flux is not yet detectable, cannot be probed with our data and is expected to be very short \citep[$\lesssim$1-2 Myr,][]{Johnson03,Corbelli17,Knutas25}.
\begin{figure}
    \centering
    \includegraphics[width=0.9\linewidth]{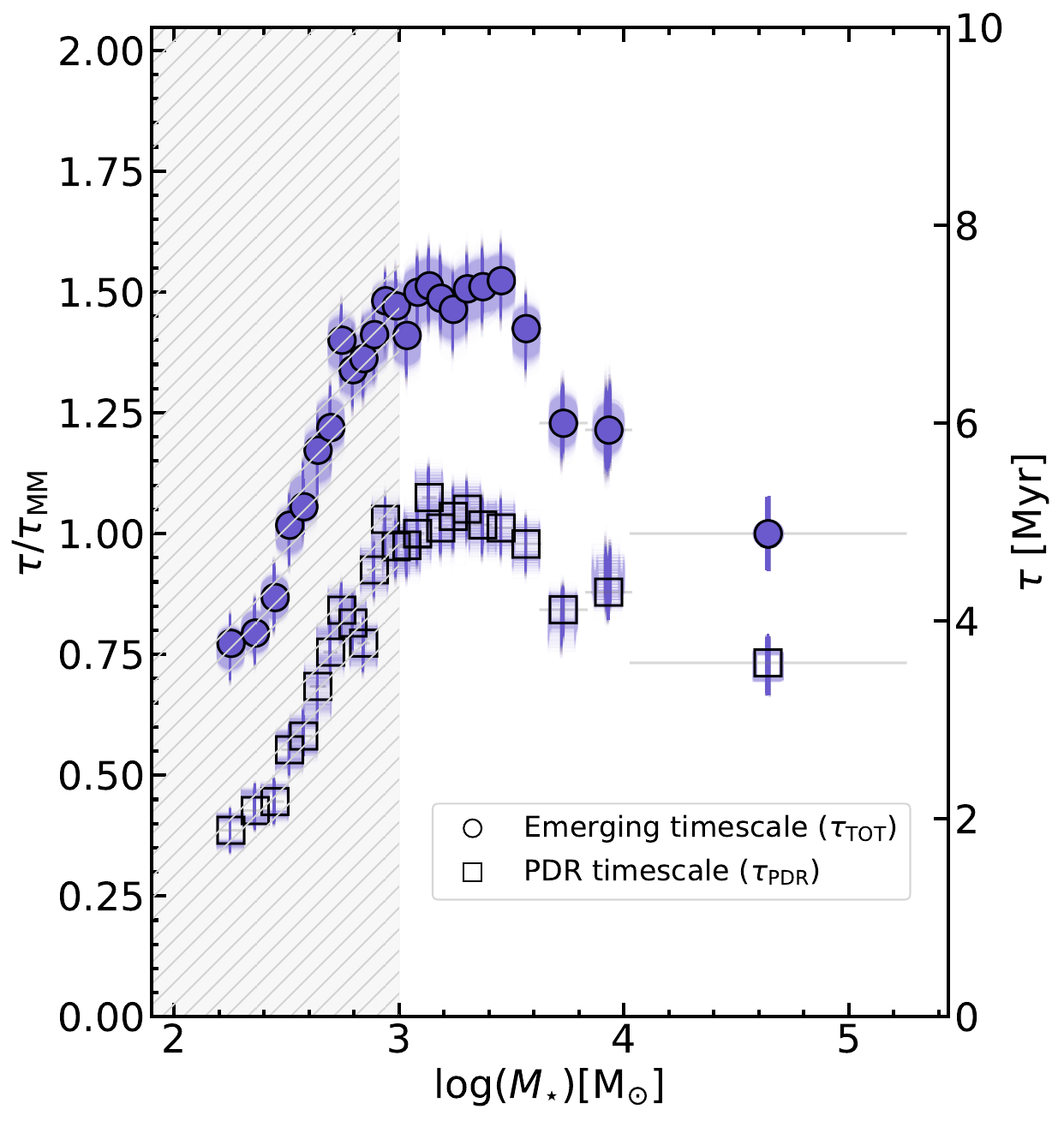}
    \caption{Emerging timescales as a function of stellar mass (log scale) for the combined population of YSCs in the four FEAST galaxies. Filled circles and open squares indicate the $\tau_{\rm TOT}$ and $\tau_{\rm PDR}$ timescales, respectively. Each bin contains the same number of objects, except for the largest two mass bins, which are split into two additional bins as described in the text; mass binning is further detailed in the Methods. On the y-axis, $\tau/\tau_{\rm MM}$ shows timescales normalized to the $\tau_{\rm TOT}$ of the highest mass bin ($\tau_{\rm MM} = 4.9$ Myr), while the secondary y-axis gives absolute values. Shadowed areas beneath the measurements show the timescales re-estimated by randomly sampling the ages of the oYSCs (see Methods), confirming the observed trends. y-error bars show Poisson uncertainties (set by the number of sources in each bin), and x-error bars indicate the bin widths. Below $10^3$ M$_{\odot}$, results are affected by completeness limits; this region is shaded in grey. }
    \label{fig:timescales_all}
\end{figure}

Figure~\ref{fig:timescales_all} shows the median of $\tau_{\rm TOT}$ (filled circles) and $\tau_{\rm PDR}$ (unfilled squares) timescales, as a function of the logarithm of stellar mass for the combined populations of YSCs ($\sim 8900$ clusters) in the four observed galaxies. Each mass bin contains the same number of objects (356). Since the number of clusters with masses above $\sim 10^4$ M$_\odot$ is small, the highest mass bin span most of the upper end of the distribution. To avoid having one excessively wide bin, we subdivided it into two additional bins of 188 clusters. This improves the temporal sampling while still ensuring a statistically significant number of clusters in each phase.
We further assessed the impact of uncertainties in the physical properties (from the {\tt CIGALE} fits) on the recovered trend by propagating errors through Monte Carlo sampling, shown as shaded areas underlying the measurements. The cluster sample is complete above $10^3$ M$_{\odot}$ \citep{Pedrini25}. The grey shaded region indicates lower masses where incompleteness affects the sample.

On the right y-axes, we plot absolute timescales from Equation~\ref{eq:total} and ~\ref{eq:pdr}, and on the left y-axes the normalized timescales to the value recovered for the $\tau_{\rm TOT}$ in the highest mass bin (blue circle corresponding to $\tau_{\rm MM}=4.9$ Myr and $\tau/\tau_{\rm MM}=1$). While the absolute timescale will somewhat depend on the maximum age assumed to select YSCs and fit eYSCs (10 Myr), the relative timescales enhances the differences as a function of cluster stellar mass. 

Above $10^3$ M$_{\odot}$ (where detection is complete at all emergence stages), we find a strong trend with stellar mass, where massive cluster $\tau_{\rm TOT}$ (filled circle) emerge  faster ($\sim 5$ Myr) than less massive $\tau_{\rm TOT}$ ($\sim 7-8$ Myr). Clusters in the low mass bins are about 1.5 times slower than the most massive clusters to complete the emerging sequence. 
Moreover, high mass clusters spend most of their emerging phase ($\sim$75\% of the time, $\sim$ 4~Myr, open squares) associated to a compact PDR traced by 3.3 $\mu$m emission, whereas low mass clusters stay associated with a compact PDR for $\sim$65\%  ($\sim$ 5~Myr) of their total emerging timescale (7-8~Myr). This implies that, after losing their PDR, low mass clusters require relatively longer time to complete their emergence than their massive counterparts (difference between circles and squares at each mass bin).




\begin{figure}
    \centering
    \includegraphics[width=0.9\linewidth]{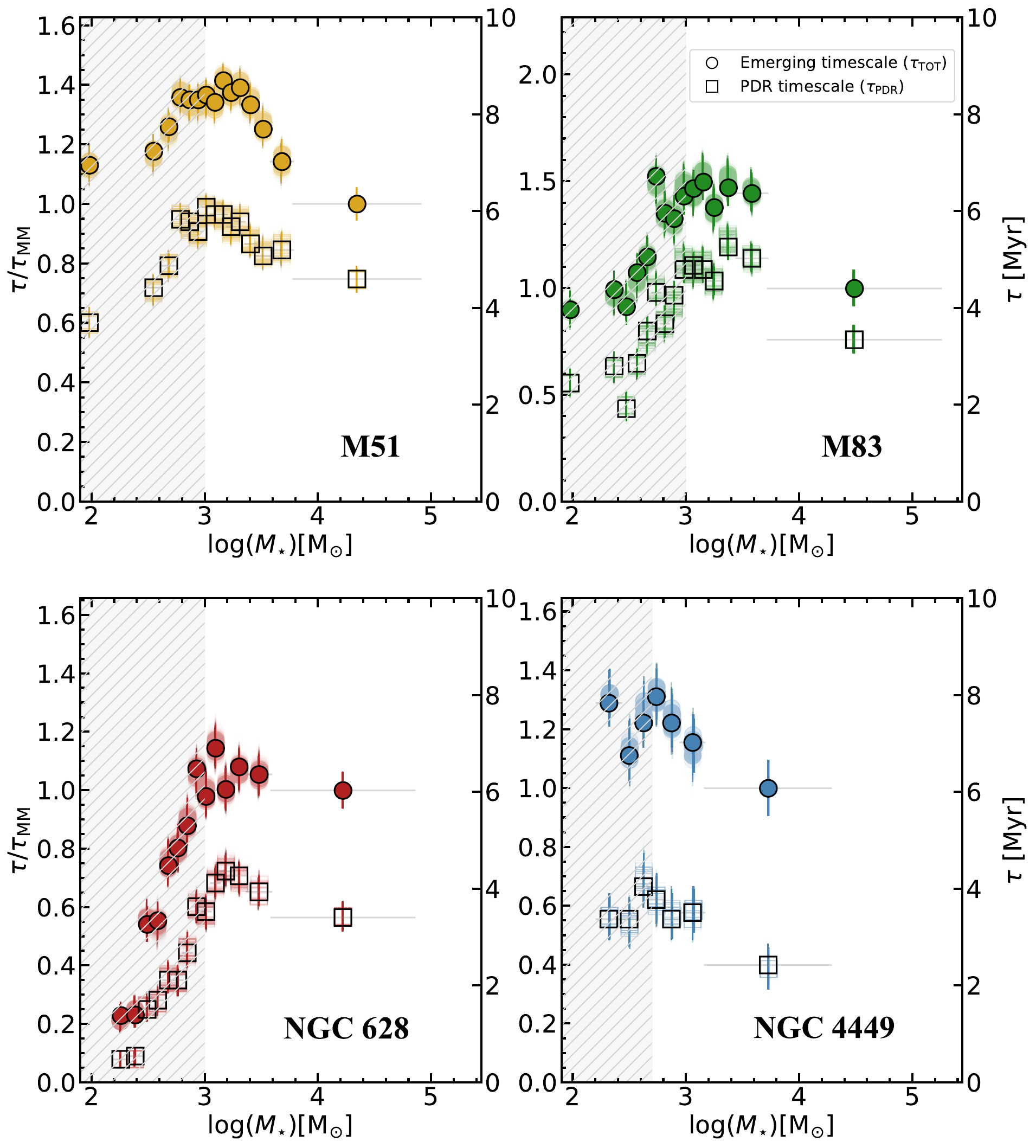}
    \caption{Emerging timescale as a function of stellar mass (log scale) for YSCs in the four FEAST galaxies individually: M51 (top left), M83 (top right), NGC 628 (bottom left), and NGC 4449 (bottom right). Symbols, binning, and shading are as in Figure~\ref{fig:timescales_all}. In NGC 4449, masses are affected by completeness limits below 10$^{2.75}$ M$_{\odot}$ \citep{Pedrini25}.}
    \label{fig:timescales_single}
\end{figure}
Figure~\ref{fig:timescales_single} presents the timescale analysis for each galaxy, confirming the general trends observed in Figure~\ref{fig:timescales_all}. However, it also reveals differences among galaxies. M51 shows the longest timescales, with emerging timescales peaking at $\sim$~9 Myr for clusters in the low mass bin, possibly affected by complex galactic dynamics due to on-going tidal interactions with a satellite galaxy \citep{Dobbs10}. M83 and NGC 628 follow more closely the overall population trend, although in NGC 628 the contrast between low and high mass bins is less pronounced, with comparable timescales across bins. Differences in the recovered emerging timescales among the three spirals might arise from variations in their cluster birth environments, gas surface densities, cloud mass spectra, and dynamical conditions (e.g., shear, bars, shocks). 
Interestingly, NGC 4449, the low metallicity target ($1/3$ Z$_{\odot}$ \cite{Pilyugin15}), exhibits much shorter $\tau_{\rm PDR}$ than the other three spirals. \cite{Gregg26} reported a deficit of PAHs in this galaxy, possibly due to the limited formation of these molecules in such environments and/or their destruction, as also confirmed by recent analyses of other low metallicity targets \citep[e.g.,][]{Lai25}. Similarly, Sabbi et al.~(subm.), using HST/FUV observations of NGC 4449, found that stars and clusters do not exhibit a UV bump at 217.5 nm in regions with intense UV radiation suggesting rapid disruption.

The results presented in this work rely on the assumption that the derived stellar mass does not increase significantly during the emergence process, since we select star clusters already associated with compact H\,\textsc{ii} regions. This assumption is reinforced by observations of YSCs in the Milky Way at similar physical scales as in this study, where their member stars present an age spread of less than 0.5 Myr, suggesting that cluster assembly occurs on short timescales \cite{Kudryavtseva12}.

However, simulations that account for stellar dynamics in emerging star clusters suggest that between 5 and 20\%  of the stellar mass may be lost during these initial stages \citep[e.g.,][]{Pfalzner13, Guszejnov22}. In Methods, we present tests investigating the impact of stellar mass loss by considering two extreme scenarios in which eYSCs lose 20\% and 50\% of their total stellar mass before they become YSCs. Even under these extreme assumptions, our findings show that the mass trend is preserved.

We additionally tested the robustness of the mass–emergence timescale relation by removing each galaxy in turn, and the trend persists.
Finally, to avoid potential biases from SED fitting, we analyzed the emerging timescales as a function of cluster luminosity in broad band NIR filters. The general trends remain unchanged when timescales are plotted against stellar continuum luminosity (Extended Data Figure~\ref{fig:lumTrends}).

\section{Discussion}
In recent years, several studies have investigated the emergence of star formation at larger physical scales ($\sim$ 25-150 pc) by analyzing the spatial de-correlation of tracers of different ISM phases \cite{Kawamura09,Corbelli17,Chevance22,Kim23,Kim25,Ramambason25}. \cite{Kim25} reported timescales for 17 galaxies of the overlapping phase between H$\alpha$ and PAH tracers (e.g., 7.7 $\mu$m), which can be thought of as a proxy for our $\tau_{\rm PDR}$. The authors found a median value of 6 $\pm$~1~Myr, slightly higher than our results for this timescale (4-5 Myr depending on the specific mass bin, Figure~\ref{fig:timescales_all}), although still within the uncertainties. At star cluster scales, other JWST studies have reported timescales equal or shorter than 5 Myr for their PDR dominated phase \citep{Linden24,Rodriguez25, Sun24, Whitmore25}, consistent with the average values found in this work. 
However, none of these studies have explored the dependence of the emerging timescales as a function of cluster stellar mass. 
In our study, we provide the first galaxy-wide, cluster-scale census of embedded to exposed phases linking emergence timescales to mass across star-forming galaxies representative of the range observed in the Local Volume. Massive clusters clear faster, implying earlier UV leakage and reduced gas replenishment time at precisely the sites that dominate a galaxy’s ionizing budget. The timescales recovered in this study confirm that pre-supernova feedback plays a pivotal role in pre-processing the gas, prior to the explosion of the first supernovae \citep{Schinnerer24}. During $\tau_{\rm PDR}$, radiation pressure drives the expansion of H\,\textsc{ii} regions \citep{McLeod21,Barnes21,DellaBruna22}, which in turn disperse the surrounding PDRs. 
Our recovered total timescales, $\tau_{\rm TOT}$, are consistent with the expected timing of supernovae, with the observed decreasing trend with mass supporting a scenario in which supernova-delay time is longer for lower mass YSCs \citep{Chevance22}.

\begin{figure}[htb!]
    \centering
    \includegraphics[width=0.496\textwidth]{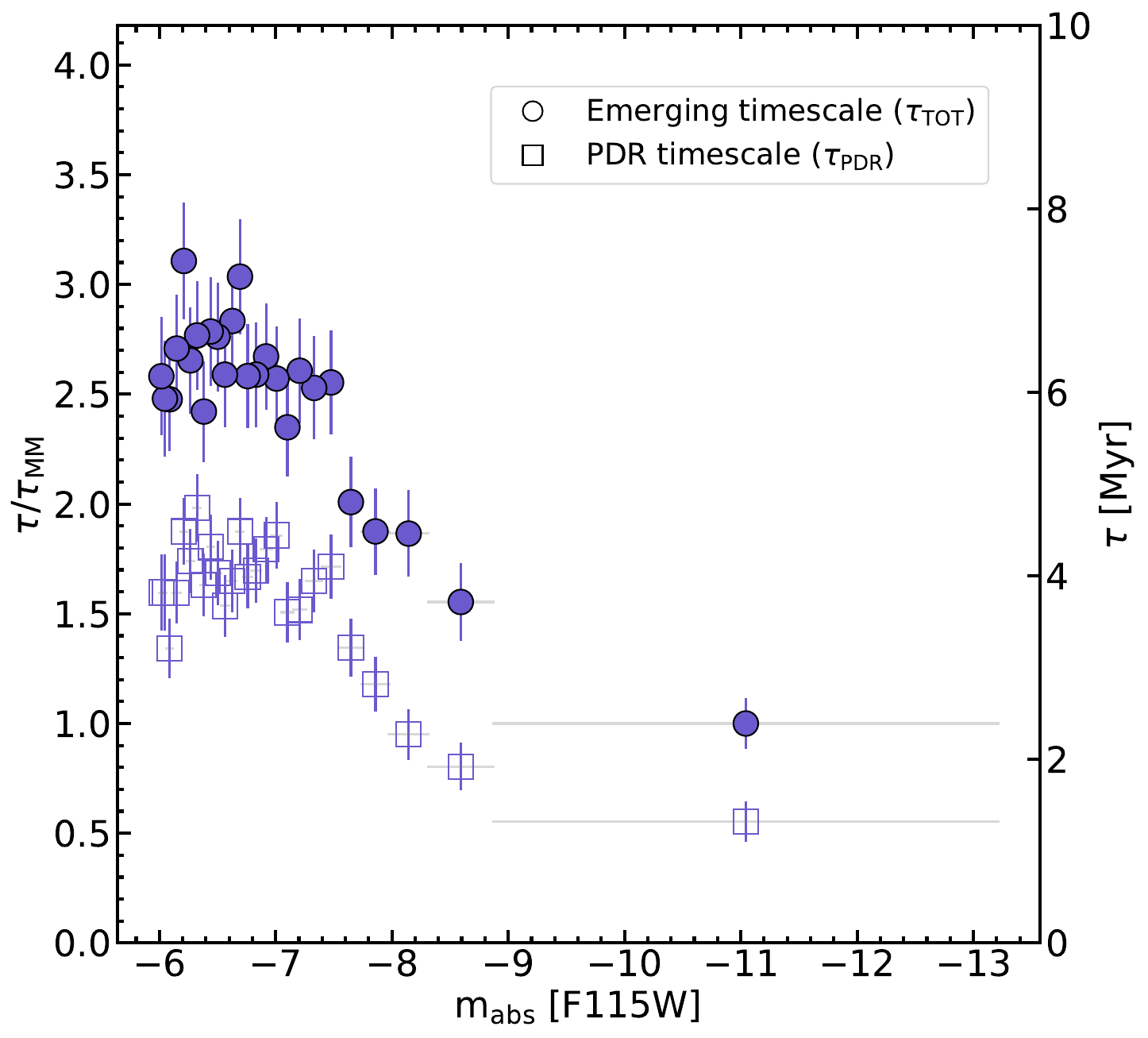}
    \hfill
    \includegraphics[width=0.496\textwidth]{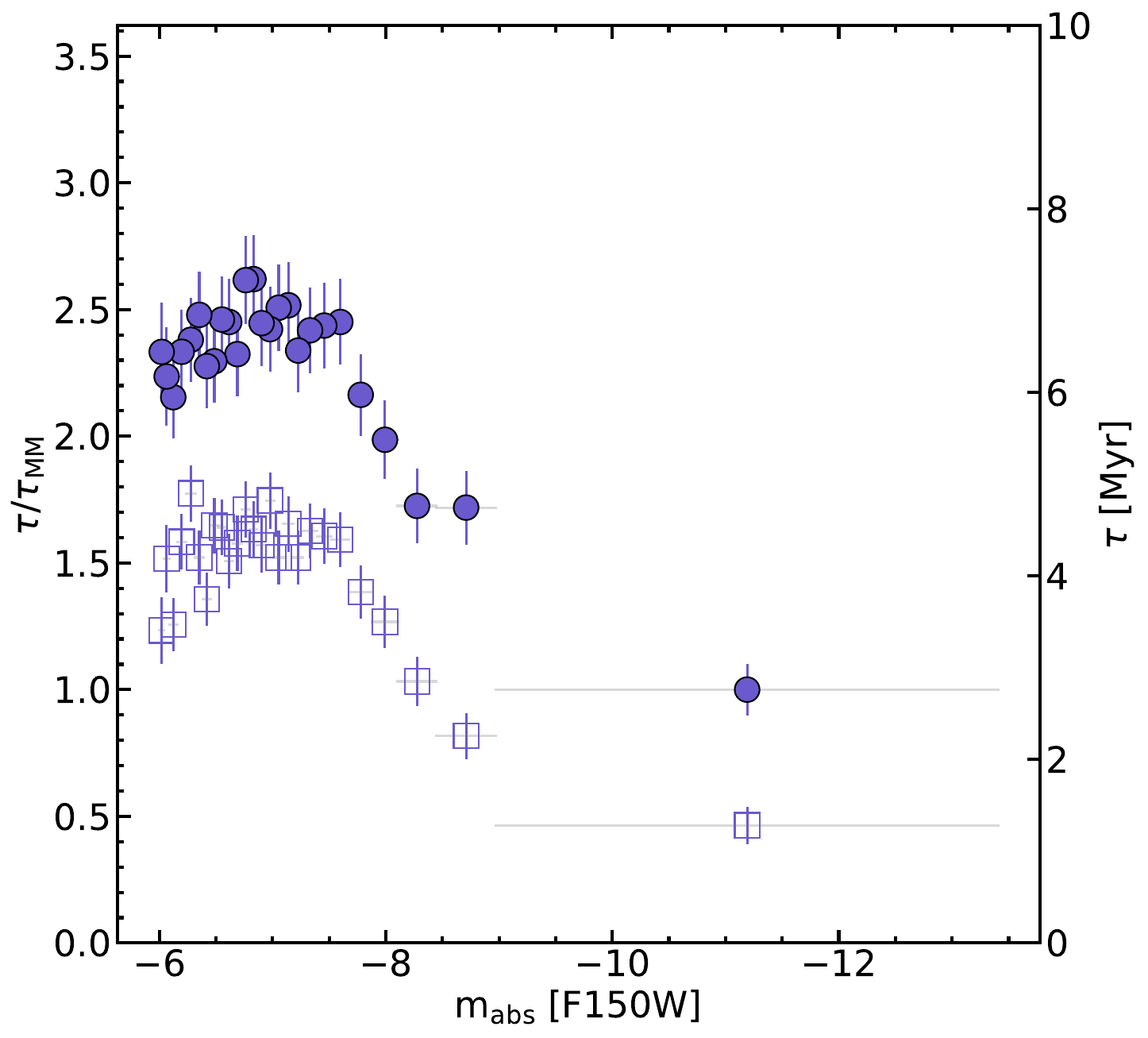}
    \\[\smallskipamount]
    \includegraphics[width=0.496\textwidth]{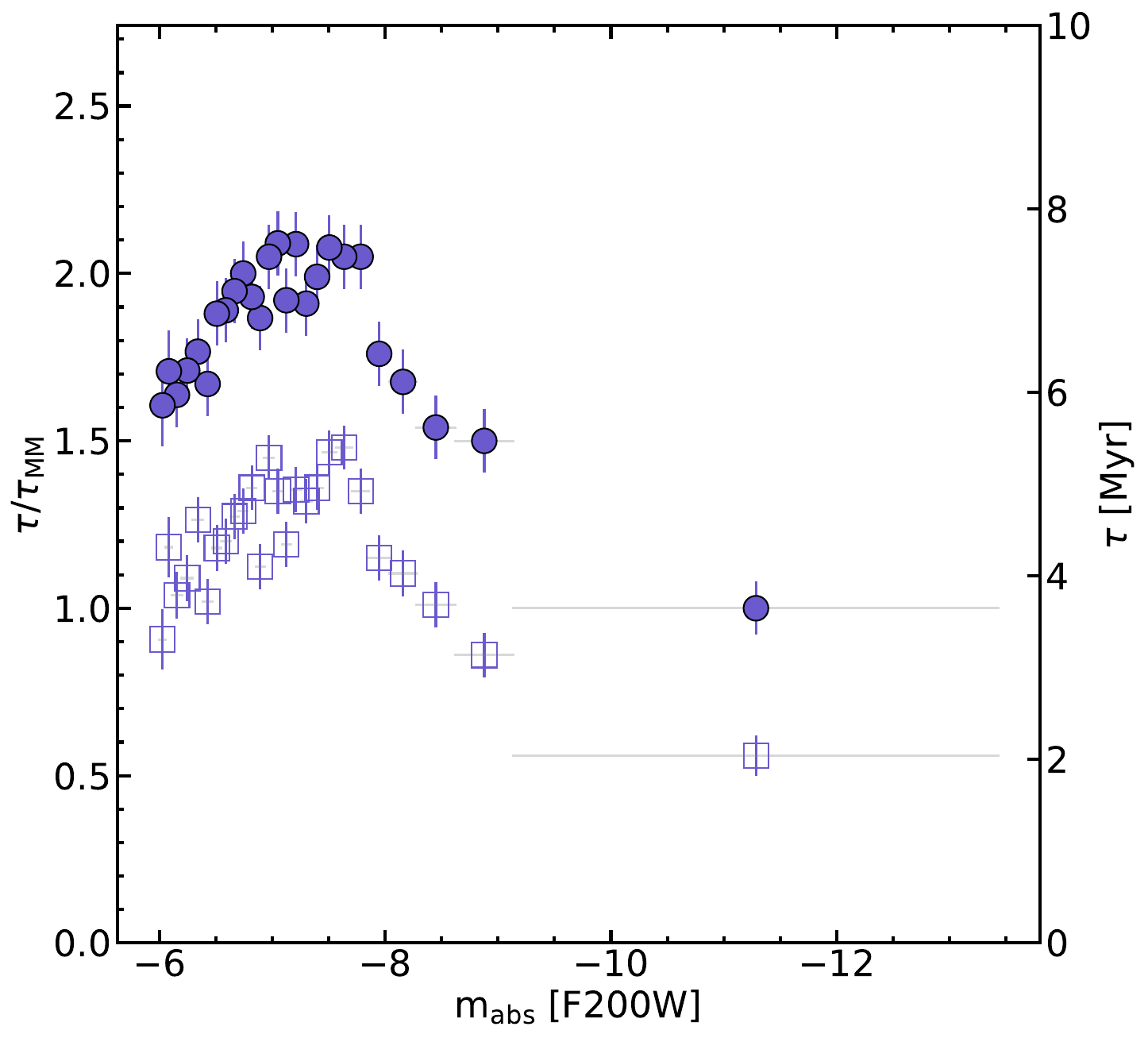}
    \hfill
    \includegraphics[width=0.496\textwidth]{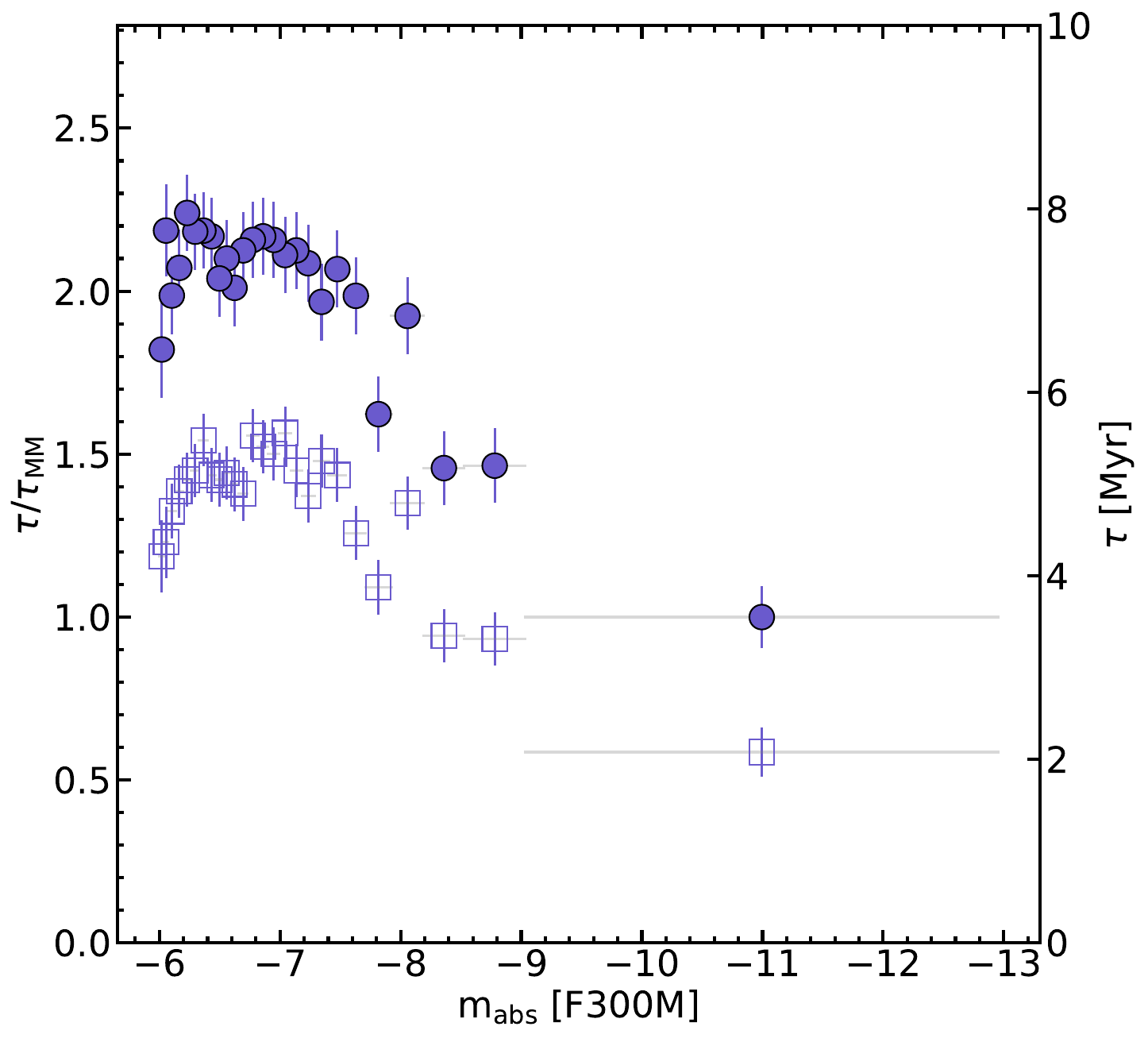}

    \caption{Emerging timescales as a function of luminosity (absolute AB magnitude) for the combined population of YSCs in the four FEAST galaxies. The four panels display four different NIRCam filters: F115W (top left), F150W (top right), F200W (bottom left), F300M (bottom right) For each filter, we applied a luminosity cut at m$_{\rm abs} > -6$. Symbols and errorbars are as in Figure~\ref{fig:timescales_all}.}
    \label{fig:lumTrends}
\end{figure}

The cluster masses analyzed in this work rely on SED fitting estimates from {\tt CIGALE}, which adopts a deterministic approach. However, stochastic initial mass function sampling effects become increasingly important at lower cluster masses \citep{Cervino04}. 
We verified this explicitly by replacing the inferred mass with integrated NIR luminosity and found that the same trends persist (Extended Data Figure~\ref{fig:lumTrends}), thus confirming that the recovered timescale dependence on cluster mass is reliable. The flattening and scatter of $\tau/\tau_{\rm MM}$ below $\sim 10^{3.5}$ M$_{\odot}$ is driven by variable number of massive stars that start to affect the total cluster luminosity and the inferred stellar mass.

State-of-the-art numerical simulations of isolated clouds reach mixed conclusions on the dependence of emerging timescales on star cluster mass. Initial conditions of the molecular gas clouds and implemented recipes for formation of stars and feedback treatment are likely the cause of these discrepancies \citep{Grudic24}. Analyzing STARFORGE simulations \citep{Grudic21}, \cite{Wainer25} report longer emerging timescales for more massive clusters if they form in more massive but less dense molecular clouds. Conversely, other studies show that massive clusters achieve higher integrated star formation efficiency and emerge faster (dissolve their natal cloud more rapidly) if they form in increasingly denser clouds \citep{Kim18, Polak24a, Menon25}. At high gas densities, radiation pressure on dust and gas becomes the main feedback mechanism responsible for emerging, allowing the expansion and then the dissolution of H\,\textsc{ii} regions \citep{Menon25}. Although the starting points of simulations and our analysis differ (our detection is dictated by the presence of ionized hydrogen and PAH emission), our results might guide simulations toward assumptions that yield more realistic outputs and calibrate star cluster feedback for sub-resolution feedback models. Extragalactic observations generally suggest an almost-flat trend between cluster size and mass \cite{Krumholz19}. Hence, by assuming that clusters in our analysis have similar sizes (in local galaxies, cluster sizes show log-normal distributions around 2-3 pc \citep{Krumholz19}), more massive star clusters are on average denser. If the higher stellar density reflects the initial density of the gas clumps from which the cluster formed, the trends we observe support the scenario in which massive star clusters do not simply form in more massive clouds, but in denser gas clumps, resulting in higher integrated star formation efficiency and faster cloud clearing timescales. 

Because the number of massive stars and the mass of the most massive star scale with cluster mass \citep[e.g.,][]{Weidner10}, it is expected that massive clusters dominate the production of ionizing photons in galaxies \citep{Stanway23}. The key question we address here is how long does it take for clusters to emerge from their birth cloud. Until now, it has been unclear whether low or high mass clusters emerge faster, and how this affects ionizing photon production and escape. 
Our findings show that massive clusters clear their natal molecular clouds more rapidly, confirming them as the dominant accessible sources of ionizing photons in galaxies. 

Additionally, our results have important implications for the theory of planet formation. Planet-forming disks around stars are influenced by their immediate environment \citep[e.g.][]{Longmore14,Winter22}. Noticeable differences are found in disk fractions near massive and dense star clusters, where UV photoevaporation and stellar encounters become dominant \citep{Stolte15,Pfalzner22}, with the most massive clusters (M $> 10^4$ M$_\odot$) showing significantly lower disk fractions compared to lower-mass clusters of the same age. Recent observations and simulations \citep[e.g.][]{Pineda20, Kuffmeier23, Gupta24, Padoan25, Haworth25} have reported prolonged accretion from surrounding dense gas into disks. The shorter emergence timescales of massive clusters have the dual effect of exposing a larger fraction of disks to external photoevaporation and removing the mass reservoir for infall more rapidly. Achieving such an interpretation on a cluster-by-cluster basis in the Milky Way is challenging given the specifics and complexities of each system. However, our statistical sample of extragalactic clusters provides a more general and statistically robust framework.
Our findings that massive clusters experience earlier decoupling between stars and gas are consistent with the observed lower disk fractions in massive clusters in the Milky Way, which would reflect in shorter timescales available for planet formation \citep{Lin23}.

\section*{Methods}

\subsection*{Observations and Data Reduction}
In this work, we used publicly available HST UV-optical observations of M51, M83, NGC 628 and NGC 4449, as well as newly collected NIR observations with JWST. The properties of these galaxies are described in \cite{Pedrini25}. Moreover, an overview of the observed HST filters and the corresponding observing programs can be found in Table~2 of \cite{Pedrini25}. 
JWST NIRCam observations were obtained from the the Mikulski Archive for Space Telescopes (MAST) as part of the FEAST program. These galaxies have been selected to span a range of distances between 4 and 10 Mpc, ensuring resolutions of a few parsec in the NIR emission coming from H\,\textsc{ii} regions and PAHs in PDRs, as well as access to large field of views covering most of the galactic disks. They also encompass a broad range of galactic environments, including nuclear starbursts and molecular rings, as well as an interacting system, a low pressure dwarf system, and spiral galaxies with diverse arm morphologies. Our targets have been observed in the following filters: F115W, F150W, F187N, F200W, F300M (F277W in the case of NGC 628), F335M, F405N, and F444W. The data have been reduced with the public pipeline (version 1.12.5 and calibration number 1169). Both HST and JWST images have been drizzled to a common scale of 0.04"/pix and aligned to the same reference system with GAIA \citep{Gaia23}. To obtain emission maps of the Pa$\alpha$-1.87$\mu$m hydrogen recombination line from the F187N filter, we subtracted the stellar continuum using the two adjacent filters F150W and F200W. The F200W also contains emission from the Pa$\alpha$ line and therefore we developed an iterative procedure to remove this additional component. Similarly, we used the F300M (or F277W in NGC 628) and the F444W to subtract stellar and dust continuum from the F335M and the F405N filters to obtain maps of the 3.3 $\mu$m PAH emission and the Br$\alpha$ recombination line. These procedures are fully described in \cite{Gregg24} and \cite{Calzetti24}. Final science-ready mosaics and continuum
subtracted maps for the FEAST galaxies will be available at \url{https://
feast-survey.github.io}. 
\paragraph*{Identification of eYSCs and oYSCs}
We identified eYSC by the presence of compact and peaked emission in the Pa$\alpha$/Br$\alpha$ emission lines and in the 3.3 $\mu$m PAH maps. To do this, we performed aperture photometry from the visually selected clusters in all the available HST and JWST filters. The adopted method for the multi band photometry is fully described in \cite{Knutas25}. We used radii of 5 pixels for the closer galaxies M83 and NGC 4449, corresponding to a physical radius of about 4.55 and 3.88 pc, respectively. For M51 and NGC 628, we used radii of 4 pixels, which corresponds to 5.8 and 7.8 parsec, respectively. Clusters with a clear and compact detection in Pa$\alpha$ and a compact 3.3 $\mu$m PAH counterpart are classified as eYSCI. On the other hand, eYSCII do not show peaked emission in the 3.3 $\mu$m PAH. We adopted a 4-pixel matching criterion between the center of the Pa$\alpha$ and the 3.3 $\mu$m PAH peaks in order to be selected as a eYSCI. As described in \citep{Knutas25}, a larger (5 px) matching criterion has been tested to verify the effect of this assumption in the distinction of eYSCI vs. eYSCII. The changes across classes were negligible (a few percent), not changing significantly the recovered timescales. oYSCs are selected in the HST/F555W filter and to be younger than 10 Myr, where cluster age is estimated from {\tt CIGALE} (see next section). The reader may find an in-depth summary of the selection process in \cite{Pedrini24,Gregg24,Knutas25,Pedrini25}. After photometry, we applied a S/N $> 3$ cut to eYSCI and eYSCII in all the NIRCam bands, in order to ensure good detection of these objects. On the other hand, oYSCs are required to be well detected in at least four HST bands. 
Additionally, a few percents of the eYSCs found a match in position to an oYSCs. We removed these objects from the eYSC catalogs and retained them only in the YSC class.

\subsection*{Derivation of physical properties and emerging timescales}
We derived physical properties of eYSC and oYSCs using the SED fitting code {\tt CIGALE}. In short, for each star cluster {\tt CIGALE} uses a deterministic approach to find the set of physical models (stellar, nebular and dust) that best fits the observations. The goodness of the fitting process is measured by the reduced chi-square ($\chi^2_{\rm red}$). The entire fitting procedure, as well as the resulting distribution of the ages and masses of eYSC, is presented in \cite{Pedrini25}. After visual inspection of the SEDs, we applied a $\chi^2_{\rm red} < 50$ selection for all the fitted clusters to rule out the presence of large fit failures in the sample. The identification criteria for eYSC detection implies that these clusters are young, with ages below 10 Myr. Their mass ranges between 10$^2$ and 10$^4$ M$_{\odot}$, with a peak around 10$^3$ M$_{\odot}$ and the median attenuation values E(B-V) vary from 0.6 mag in NGC 4449 and NGC 628 to 0.8 mag in M51 and M83. 
The authors of this paper recommend caution in relying on absolute values of age, due to the lack of available models that account for all the components necessary to fit YSCs in the NIR, such as stochasticity of the initial mass function sampling, pre-main-sequence stars and hot dust.
For this reason, in this paper we estimated emerging timescales by leveraging a statistical approach that has been previously explored in the literature and does not involve single cluster age estimations \citep{Whitmore23, Knutas25}. We can count the fraction of clusters in the identified evolutionary phases (see Fig.~\ref{fig:NIRcolors}) and define the emerging timescale $\tau_{\rm TOT}$ introduced in Equation~\ref{eq:total}.
We adopted 10 Myr as a star-formation timescale, since the eYSC are fitted up to 10 Myr, due to the presence of ionized gas emission, and the oYSCs are selected to be younger than 10 Myr. We notice that our trends do not change if we use a different limit for the oYSCs ages (e.g., 7 Myr or 15 Myr). Below $\sim 10^3$ M$_{\odot}$ (shaded area), our results are sensitive to completeness limits and the positive trends for the emerging timescales might be due to more severe incompleteness in individual cluster catalogs.
Analogously, we can define a timescale necessary for eYSCI to become eYSCII, i.e., the timescale for cluster to lose their compact PDR traced by 3.3 $\mu$m PAH emission. This is the $\tau_{\rm PDR}$, as defined in Equation~\ref{eq:pdr}.
With these definitions, we note that uncertainties on the recovered ages with {\tt CIGALE} may lead to an incorrect estimation of the number of oYSCs younger than 10 Myr, which in turn affects the recovered timescales. To assess this effect, we re-estimated the emerging timescales with a Monte Carlo sampling, drawing ages of oYSCs from normal distributions centered on the original values with widths set by their age uncertainties. We repeated this procedure 1000 times, and the resulting spread is shown as the shaded region underlying the measurements in Figure~\ref{fig:timescales_all} and ~\ref{fig:timescales_single}.

The recovered emerging timescales are then illustrated as a function of stellar mass bins, which were chosen such that each bin contains the same number of clusters. For the combined dataset (Figure~\ref{fig:timescales_all}), each bin contains 356 star clusters. Since only about 3\% of the clusters have masses higher than 10$^4$ M$_{\odot}$, we further divided the most massive bin into two sub-bins of 188 sources each, as described in the Results. For the analysis of individual galaxies, each bin contains 199, 171, 165, and 74 clusters for M51, M83, NGC 628, and NGC 4449, respectively. For the single-galaxy cases, we did not apply any splitting to the high-mass bins.

\paragraph*{Impact of stellar mass loss during the emerging process}
In this analysis, by comparing the counts of eYSCs and oYSCs within the same mass bins to derive emerging timescales, we implicitly assume that the stellar mass remains constant during the emergence to the optical phase. This assumption warrants further investigation. Mass losses due to stellar evolution and ejection of stars due to strong dynamical interactions \citep[e.g.,][]{Oh12,Oh16,Pfalzner13b,Renzo19,Guszejnov22} may be able to reduce the final mass of oYSCs by a non-negligible amount.

\begin{figure}[htb!]
    \centering
    \includegraphics[width=0.496\textwidth]{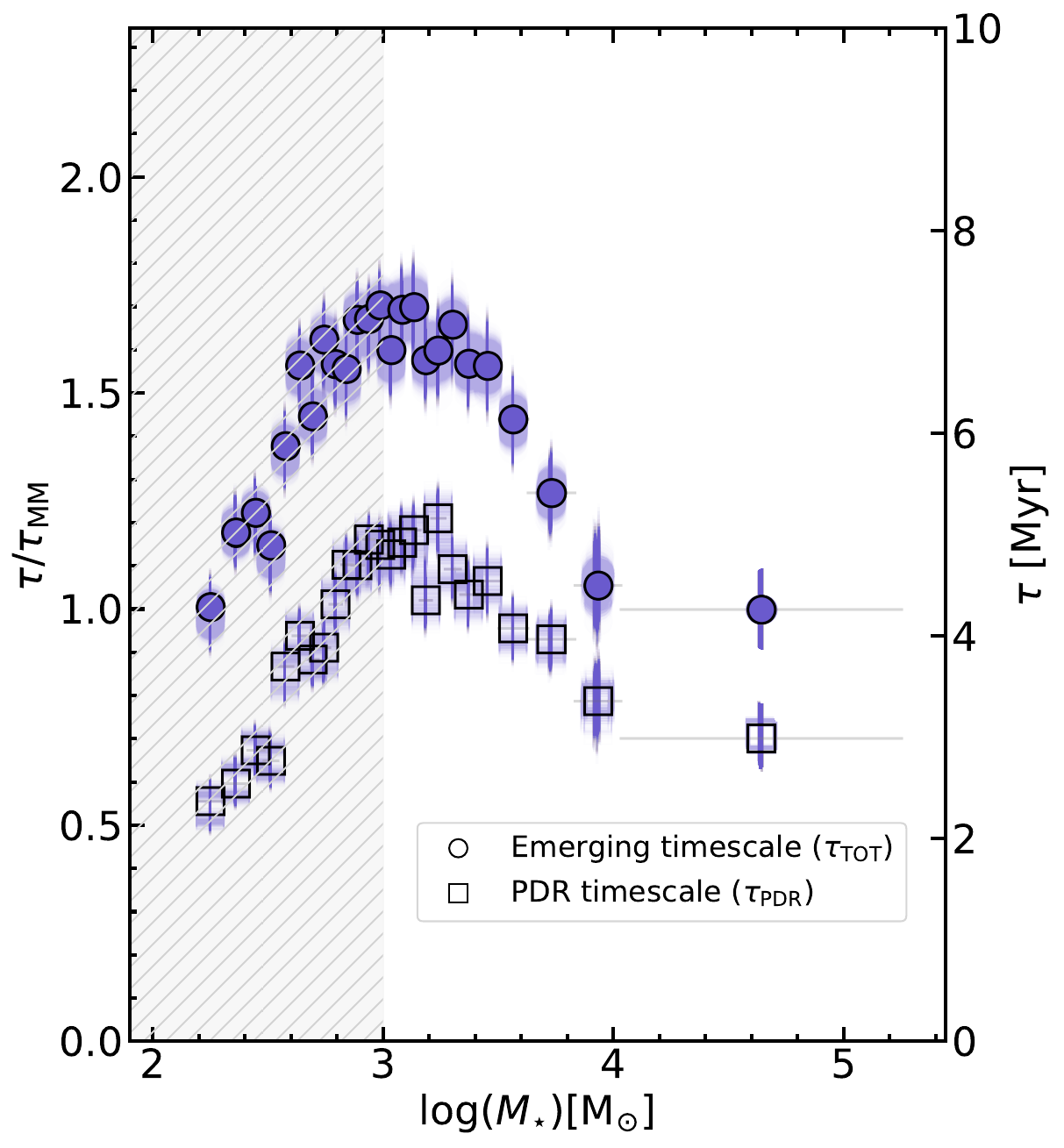}
    \hfill
    \includegraphics[width=0.496\textwidth]{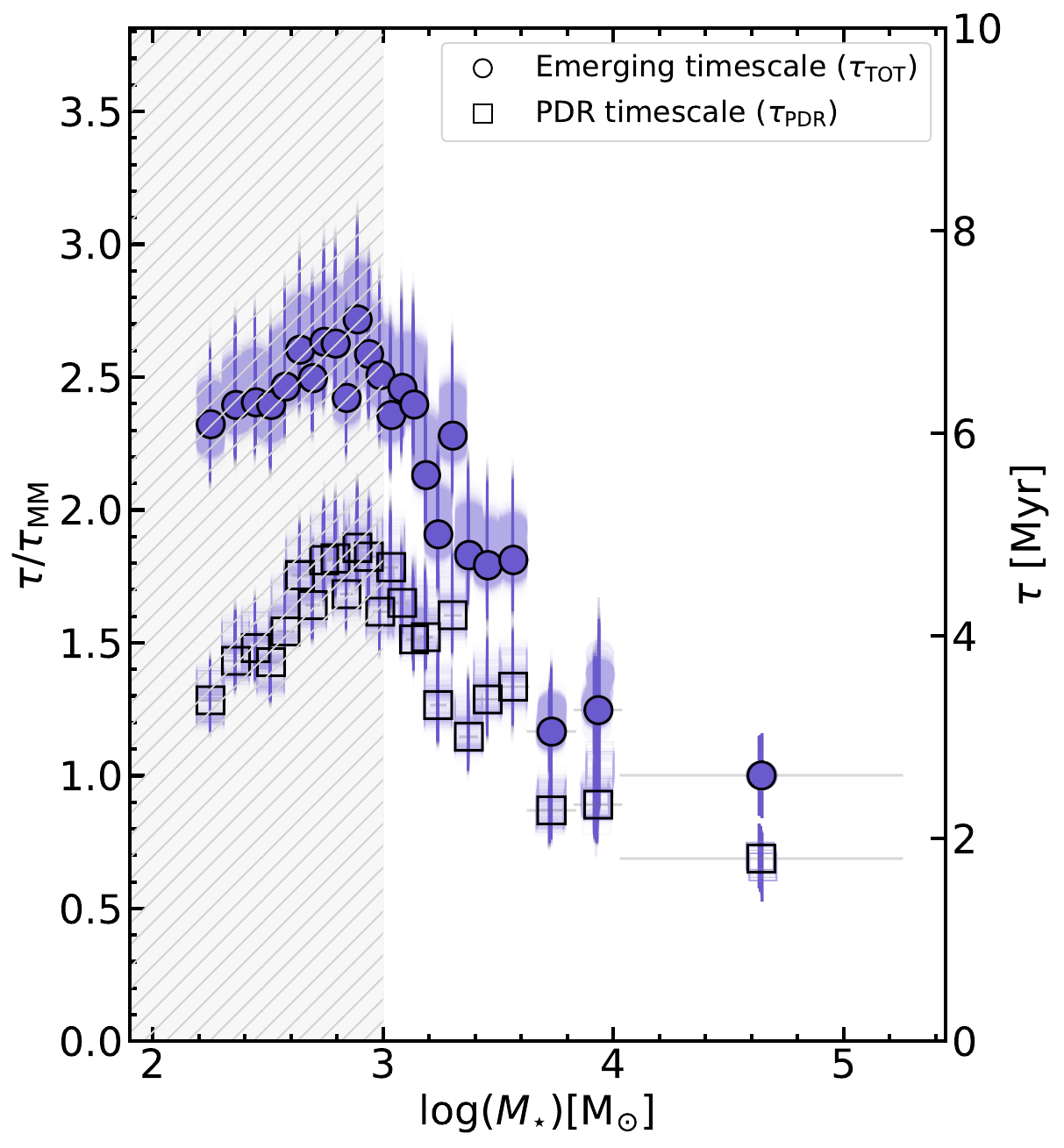}
    
    \caption{Emerging timescales as a function of {\bf stellar} mass (log scale) for the combined population of YSCs in the four FEAST galaxies, accounting for stellar mass loss. The left (right) panel shows the scenario in which 20\% (50\%) of the stellar mass is lost during the emerging process from the eYSC to the oYSC phase, as described in the text. Symbols and errorbars are as in Figure~\ref{fig:timescales_all}.}
    \label{fig:massLoss}
\end{figure}

We test extreme scenarios in which mass is removed instantaneously, and eYSCs lose between 20\% and 50\% of their total stellar mass during their emergence into oYSCs. To perform this test, we repeated the analysis shown in Fig.~\ref{fig:timescales_all} after multiplying eYSC stellar masses by factors of $4/5$ (20\% mass loss) and $1/2$ (50\% mass loss). With this approach, eYSC with stellar mass $m_*$ are assigned to the same mass bins as oYSCs with masses $4/5 \, m_*$ and $1/2 \, m_*$, respectively, thereby mimicking the effect of mass loss. We present the results of this analysis for the combined population of YSCs in Extended Data Figure~\ref{fig:massLoss}, where the left and right panels correspond to mass losses of $4/5$ and $1/2$, respectively. Under these assumptions, more massive clusters move faster toward short emerging timescales, however, the overall trend with cluster stellar mass is preserved.

\section*{Extended Data}


\begin{table}[htb!]
\begin{tabular}{c|c c c c}

     & M51 & M83 & NGC 628 & NGC 4449 \\
    \hline
    eYSCI & 1589 & 1001 & 636 & 211 \\
    eYSCII & 768 & 429 & 512 & 229 \\
    oYSCs & 664 & 1201 & 1495 & 162 \\
    \hline

\end{tabular}
\caption{Number of eYSCI, eYSCII and oYSCs identified in four galaxies from the FEAST sample. We note that the number of eYSCs differs from that reported in \cite{Pedrini25}, due to the additional $ \chi^2_{\rm red}$ cut ($\chi^2_{\rm red} < 50$) and correction for duplicates; systems that are identified both as eYSCs and oYSCs are removed from the eYSC classes and retained as oYSCs.}
\label{tab:numbers}
\end{table}

\section*{Data availability}
The data were obtained from the MAST at the Space Telescope Science Institute. The specific observations analyzed can be accessed via \url{http://dx.doi.org/10.17909/f4vm-c771}. All our data products are available at MAST as a High Level Science Product via \url{https://doi.org/10.17909/6dc1-9h53} and DOI: 10.17909/6dc1-9h53 and on the FEAST webpage \url{https://feast-survey.github.io/}.

\section*{Code availability}
This work made use of Astropy, a community-developed core Python package and an ecosystem of tools and resources for astronomy \citep{astropy:2013}, NumPy \citep{numpy}, SciPy \citep{scipy}, Matplotlib \citep{matplotlib}, Pandas \citep{pandas}. SED fits of star clusters are performed with the publicly available code {\tt CIGALE} \citep{Boquien19}. 

\section*{Acknowledgments}
This project is based in part on observations made with the NASA/ESA/CSA James Webb Space Telescope, which is operated by the Association of Universities for Research in Astronomy, Inc., under NASA contract NAS 5-03127. These observations are associated with program \# 1783. Support for program \# 1783 was provided by NASA through a grant from the Space Telescope Science Institute, which is operated by the Association of Universities for Research in Astronomy, Inc., under NASA contract NAS 5-03127. 
A.A. and A.P. acknowledge support from the Swedish National Space Agency (SNSA) through the grant 2021- 00108. A.A. and H.F.V. acknowledges support from SNSA 2023-00260. A.B. acknowledges support from the Swedish National Space Agency (2022-00154).
M.R.K. acknowledges support from the Australian Research Council through Laureate Fellowship FL220100020. E.P.A. acknowledges support from NASA ATP grant 80NSSC24K0935. A.D.C. acknowledges the support from a Royal Society University Research Fellowship (URF/R1/19160 and URF/R/241028). T.J.H. acknowledges a Dorothy Hodgkin Fellowship, UKRI guaranteed funding for a Horizon Europe ERC consolidator grant (EP/Y024710/1) and and UKRI/STFC grant ST/X000931/1. N.L. was supported by a Gliese Fellowship at the Zentrum f\"ur Astronomie,
Universit\"at Heidelberg, Germany.

\section*{Author contributions}

A.P. designed and led the data analysis, discussion of the results, and wrote the manuscript. A.A. is the PI of the FEAST program. A.A, D.C., and A.B. contributed in designing the data analysis, the scientific discussion and the drafting of the manuscript. T.H. contributed to the discussion on the implications of our results for planet formation theories. B.G.E., M.R.K, H.F.V., E.P.A., A.A.A were deeply involved in the discussion and scientific interpretations of the results. S.L. designed the cluster SED fitting with {\tt CIGALE}. B.G. produced the continuum subtracted maps used in this work. V.B. performed the data reduction used in this work. J.E.R. designed the FEAST pipeline for cluster extraction and photometry. G.B., M.C., A.D.C., K.G., N.L., T.S.Y.L., D.L., M.M., G.Ö, E.S., L.J.S., M.T. and all coauthors contributed to writing and refining the manuscript, and participated in the discussion.

\section*{Competing interests}
The authors declare no competing interests.


\end{document}